\documentstyle[11pt,newpasp,twoside,epsf]{article}

\def\edcomment#1{\iffalse\marginpar{\raggedright\sl#1\/}\else\relax\fi}
\reversemarginpar

\begin{document}

\title{EXPLORING THE EARLY UNIVERSE WITH XMM-NEWTON}

\author{I. LEHMANN, G. HASINGER}
\affil{ASTROPHYSIKALISCHES INSTITUT POTSDAM, AN DER STERNWARTE 16, D-14482 POTSDAM, GERMANY}

\author{S.S. MURRAY}
\affil{HARVARD-SMITHSONIAN CENTER FOR ASTROPHYSICS, 60 GARDEN STREET, CAMBRIDGE, MA 02138, USA}

\author{M. SCHMIDT}
\affil{CALIFORNIA INSTITUTE OF TECHNOLOGY, PASADENA, CA 91125, USA}

\begin{abstract}
The question of the X-ray background has been largely settled over the
last 10 years using the {\em ROSAT} satellite. About 70-80 \% of the soft X-ray
background was resolved into discrete sources, which are mainly X-ray and
optically unobscured AGNs (Quasars, Sy1). Furthermore we found already 
several obscured AGNs in deep {\em ROSAT} surveys, which are predicted by the
population synthesis models for the X-ray background, based on the 
unified AGN schemes. However, deep X-ray surveys using {\em Chandra} and {\em XMM-Newton} allow to test the reliability of these models, which seem to be
still far from unique. Recent observations with {\em Chandra} and {\em XMM-Newton} have resolved most of the hard X-ray background into discrete sources.
The first deep {\em XMM-Newton} survey was performed in the Lockman Hole
region, one of the sky areas best studied over a wide wavelength range.
About 100 ksec good exposure time have been accumulated with the EPIC 
camera during the performance verification phase. We reach a limiting flux
of 0.31, 1.4 and 2.4 $\times$ 10$^{-15}$ erg cm$^{-2}$ s$^{-1}$ in the 0.5-2, 2-10 and 5-10 keV energy bands. 
A significant number of X-ray sources show hard, probably intrinsically 
absorbed X-ray spectra. We discuss the X-ray, optical and infrared properties
of these sources in comparison with the predictions of the recent
X-ray background models.
\end{abstract}

\section{Introduction}

The X-ray backgound has been a matter of intense study since its discovery by
Giacconi et al. (1962). Several deep X-ray surveys using {\em ROSAT, BeppoSAX and ASCA} (Hasinger et al. 1998, Ueda et al. 1998, Cagnoni et al. 1998, Giommi et al. 2000, Ishisaki et al. 2001) and recently {\em Chandra} (Mushotzky et al. 2000, Giacconi et al. 2001, Hornschemeier et al. 2001, Brandt et al. 2001) and {\em XMM-Newton} (Hasinger et al. 2001a) have resolved most of the 0.5-10 keV X-ray background. We have performed the deepest observations with the {\em ROSAT} satellite in the Lockman Hole field, a region with extremely low absorption due to interstellar hydrogen. About 1.4 Msec {\em ROSAT} PSPC and HRI exposures were accumulated 
for the ROSAT Deep Survey (RDS; Hasinger et al. 1998) and the Ultra Deep Survey (UDS; Lehmann et al. 2001). We reached a source density of $\sim$1000 per deg$^{2}$ at a limiting flux of 10$^{-15}$ erg cm$^{-2}$ s$^{-1}$ in the 0.5-2.0 keV band, where 70-80\% of the soft X-ray background has been resolved into discrete sources.

The optical identification of the RDS has shown that the majority of sources
($\sim$80\%) are AGNs, mainly X-ray unobscured type 1 AGNs with broad emission lines
in their optical spectral (Schmidt et al. 1998). Nevertheless there is already a significant fraction of probably instrinsically absorbed AGNs (type 2) in the UDS, where the optical spectra show narrow emission lines and/or continuum emission from the host galaxy (Lehmann et al. 2001). 
In general, we have found a larger surface density of AGNs in the X-ray than observed in any other wavelength band. The combination of the ROSAT Deeps Surveys with several shallower and larger area  ROSAT surveys has led to the marginal evidence of a constant space density of X-ray selected AGNs at redshifts $2<z<4$ (Miyaji et al. 2000) in contrast to the decrease of the space density (at $z>2.7$) of radio and optically selected QSOs (Schmidt et al. 1995, Shaver et al. 1999, Fan et al. 2000).

The cosmic X-ray background is largely due to the accretion onto supermasssive black holes, integrated over cosmic time. The population synthesis
models, based on the unified scheme of AGN, predict a large number of heavily  absorbed sources to explain the characteristic hard spectrum of the X-ray background (see Comastri et al. 1995, Gilli et al. 2001). About 80-90\% of their light produced by accretion will be absorbed by gas and dust (Fabian et al. 1998).
Deep hard X-ray surveys with {\em Chandra} and {\em XMM-Newton} enable us to study the population of faint heavily absorbed X-ray sources and their impact to the X-ray background models.

\section{{\em XMM-Newton/Chandra} Deep Survey in the Lockman Hole}

The Lockman Hole field has been observed with the {\em XMM-Newton} observatory during the performance verification phase (see Figure 1). A total of 100 ksec good data, centered on the same sky position as the {\em ROSAT HRI} pointing, have been accumulated with the European Photon Imaging Camera (EPIC) reaching minimum fluxes of 0.31, 1.4 and 2.4 $\times$ 10$^{-15}$ erg cm$^{-2}$ s$^{-1}$ in the 0.5-2, 2-10 and 5-10 keV energy bands. Within an off-axis angle of 10 arcmin we have detected 148, 112 and 61 sources, respectively. 
In the 5-10 keV energy band we reached a similar sensitivity compared to the 1Msec {\em Chandra} Deep Field South observation (see Tozzi et al. 2002) and have resolved $\sim$60 \% of the very hard X-ray background (Hasinger et al. 2001a). This is about a factor of 20 more sensitive than the previous {\em BeppoSAX} observations.  

The advantage of {\em XMM-Newton} compared to {\em Chandra} is its wide energy band and its unprecedented sensitivity in the hard band, which is well suited to study very faint intrinsically absorbed X-ray sources. 
\pagebreak
\begin{figure}[htb]
\caption{X-ray colour image of the combined and exposure corrected {\em XMM-Newton} PN and MOS images of the Lockman Hole field (top). The field size is about 30 $\times$ 30 arcmin. The red, green and blue colours refer to sources in the 0.5-2.0 (soft), 2.0-4.5 (hard) and 4.5-10.0 keV (ultra hard) energy bands. 300 ksec {\em Chandra} HRC image in the 0.5-7.0 keV energy band (PI: S.S. Murray) of the same region (bottom).}
\plotfiddle{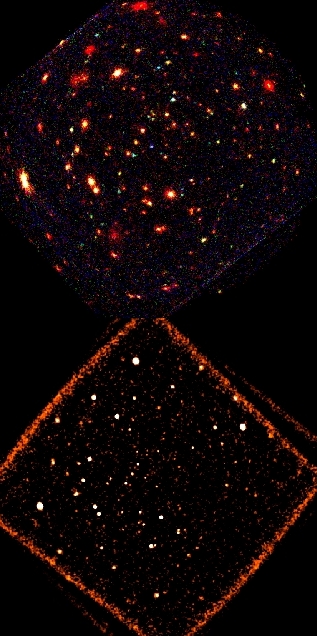}{17.0cm}{0.0}{80}{82}{-115.0}{0.0}
\end{figure}
\pagebreak

The unique image quality of the  {\em Chandra} satellite allows to take extremely deep observations (e.g., the 1Msec Hubble Deep Field-North/{\em Chandra} Deep Field-South surveys, see C. Norman's and A. Hornschemeier's contributions to these proceedings), which are not affected by source confusion. 

A deep observation of the Lockman Hole has been performed with the HRC detector
onboard the {\em Chandra} satellite. A total of 300 ksec exposure time has been
accumulated in the 0.5-7.0 keV energy band (Murray et al. 2002). 
The {\em Chandra} HRC image, as shown in Figure 1, reaches a similar flux limit in the 0.5-2.0 keV band compared to the {\em XMM-Newton} pointing. The combination of {\em XMM-Newton} and {\em Chandra} observations provides us a sufficient number of X-ray photons to analyse the X-ray properties even of faint X-ray sources and very accurate positions to identify the optical/near-infrared counterparts of these sources. 

The new {\em XMM-Newton} sources, previously not detected in the UDS, 
are fainter and typically harder than the {\em ROSAT} sources. The X-ray spectral diagnostic diagrams based on the hardness ratios derived from four different energy bands (0.2-0.5, 0.5-2.0, 2.0-4.5 and 4.5-10.0 keV) have been a first tool
to determine the possible nature of these sources (see Hasinger et al. 2001a). The new sources occupy nearly the same region as the previously known {\em ROSAT} type 2 AGNs. The majority of the new {\em XMM-Newton} sources are probably intrinsically absorbed sources, which is in good agreement with the predictions
from the X-ray background models.

About 25\% of the new sources with sufficient number of photons ($>50$) show very hard spectra, which can be described by a powerlaw plus significant intrinsic absorption log N$_{H}>22$ (V. Mainieri, priv. comm.). Deeper {\em XMM-Newton} exposures are necessary to determine the amount of intrinsic absorption in the very faint sources. Figure 2 shows two {\em XMM-Newton} sources with moderate and strong intrinsic absorption classified as type 2 AGNs. In general, the intrinsic absorption of type 2 AGN ranges from N$_{H}=10^{21}$ to $10^{24}$ cm$^{-2}$.

\begin{figure}
\caption{The {\em XMM-Newton} spectra of two type 2 AGNs are fitted with a powerlaw spectrum plus intrinsical absorption with $\Gamma=1.9$, N$_{H}=2.1 \times 10^{21}$ cm$^{-2}$, $z=0.205$ (left), and with $\Gamma=1.6$, N$_{H}=2.4 \times 10^{22}$ cm$^{-2}$, $z=0.708$ (right).}
\plotfiddle{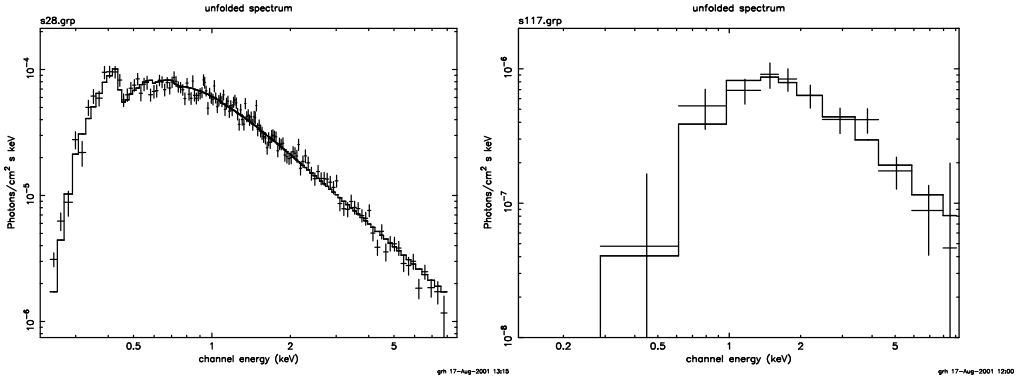}{5cm}{0.0}{26}{26}{-185.0}{0.0}
\end{figure}

\begin{figure}
\caption{Keck LRIS spectra of a narrow emission line AGN (type 2, z$=$0.708) and a type 2 QSO candidate showing narrow L$_{\alpha}$, He II and C III$]$ emission lines (FWHM$<$1200 km s$^{-1}$) at z$=$3.240.}
\plotfiddle{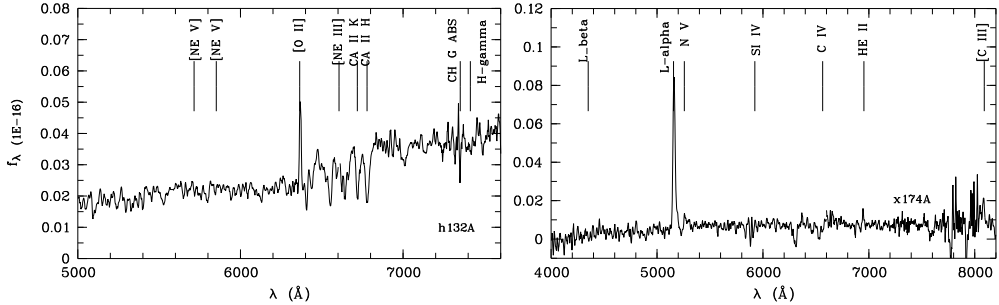}{3.2cm}{0.0}{32}{32}{-185.0}{-20.0}
\end{figure}

\section{Spectroscopic Idenification of new faint XMM-Newton sources}

The spectroscopic identification of faint X-ray sources is still the
key to understand the nature of these sources.
From the deep {\em XMM-Newton} pointing we have selected a soft, hard and ultra hard sample containing 106, 69 and 34 X-ray sources with fluxes larger than 3.8 $\times 10^{-16}$, 2.0 $\times 10^{-15}$ and 3.2 $\times 10^{-15}$ erg cm$^{-2}$ s$^{-1}$ in the 0.5-2, 2-4.5 and 4.5-10 keV energy bands. The sources are at off-axis angles below 10 arcmin, corresponding to a solid angle of 0.087 deg$^{-2}$. 
The majority of hard (57) and ultra hard (27) sources has been detected in the soft band.
 The soft sample includes three times fainter 
X-ray sources than previously detected with {\em ROSAT} in the same sky region.

\begin{table}[b]
\centering
\caption{Identification status of the {\em XMM-Newton} Deep Survey.}
\vspace{0.3cm}
\begin{tabular}{llll}
               & Soft sample & Hard sample & Ultra hard sample  \\\noalign{\smallskip} \hline \noalign{\smallskip}
AGN type 1     & 34          &   28        &   13               \\
AGN type 2    & 20          &   19        &   14               \\
AGN type 2/Galaxy& 5           &   1         &   -                \\
Galaxy            & 3           &   -         &   -                \\
Group/Cluster  & 2           &   -         &   -                \\
Star           & 1           &   -         &   -                \\
No ID, (but ERO) & 41 (15)     &   21 (10)   &   7 (3) \\\noalign{\smallskip} \hline \noalign{\smallskip}
Completeness   & 61 \%       &   70 \%     &   79 \%             \\  
\end{tabular}
\end{table}

The optical counterparts for $\sim$60 X-ray sources are already known from
the spectroscopic identification of the UDS sample (Lehmann et al. 2001).
Among them are the most distant X-ray selected quasar at $z=4.45$ found to date (Schneider et al. 1998) and one of the highest redshift, probably merging cluster of galaxies at $z=1.26$ (Hasinger et al. 1999, Thompson et al. 2001, Hashimoto et al. 2001).
About 25 new {\em XMM-Newton} sources have been identified using low-resolution
spectra taken with the LRIS instrument at the Keck II telescope in March 2001.
Surprisingly, we have found only a few new broad emission line AGNs (type 1).

The optical spectra of most new sources show narrow emission
lines and/or only galaxy-like continuum emission at redshifts $z<1.0$ (see Figure 3). In the case that high ionisation emission lines like $[$Ne~V$]$ $\lambda3426$ are absent we see no sign for AGN activity in the optical spectrum. The high X-ray luminosity (L$_{X}>10^{43}$ erg s$^{-1}$) and/or the strong intrinsic absorption (log N$_{H}>22.0$) reveal the type 2 AGN in these sources.  
Three new sources showing typical galaxy spectra, have been detected only
in the 0.5-2.0 keV band. Due to their relatively low X-ray luminosities (log L$_{X}<42.0$) and their soft X-ray spectra (no indication for intrinsic absorption) we classify them as normal galaxies. 

\begin{figure}
\caption{X-ray flux in the 0.5-2.0 keV energy band versus $R$ magnitude for all objects from the {\em ROSAT} Deep Surveys (RDS, UDS; Schmidt et al. 1998, Lehmann et al. 2000, 2001) and new {\em XMM-Newton} objects (Hasinger et al. 2001a) in the Lockman Hole field (left). ROSAT sources are shown with small symbols. The filled circles indicate type 1 AGNs, small open circles are type 2 AGNs. 
Open squares are groups and clusters of galaxies. Asterisks are stars. The crosses mark spectroscopically unidentified {\em XMM/ROSAT} sources. 
The large circles are recently identified type 1 (filled) and type 2 (open) AGNs of mostly faint {\em XMM-Newton} sources. 
The left dotted line gives the typical ratio for AGNs ($f_{X}/f_{opt}=1$), the right dotted line is plotted for three optical magnitudes fainter objects.  $R-K'$ colour versus $R$ magnitude for the same sources (right). All X-ray sources not covered by our $K'$ band survey are plotted at $R-K'=0$. }
\plotfiddle{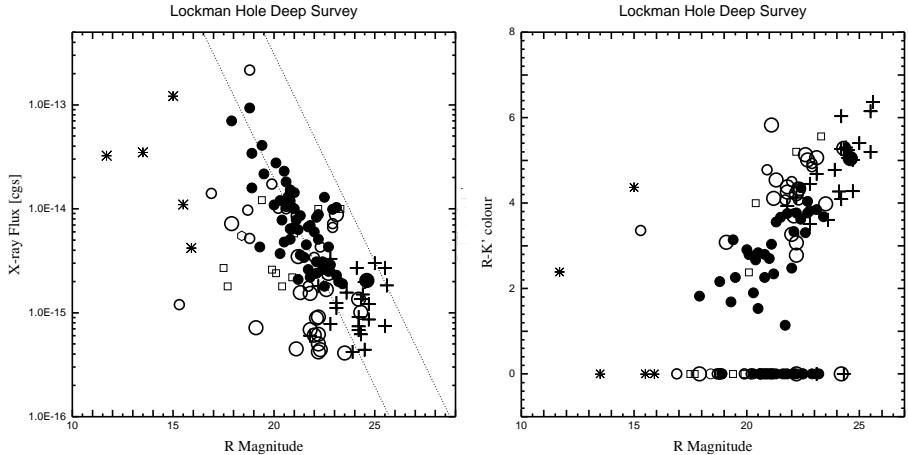}{5.3cm}{0.0}{47}{47}{-170.0}{-15.0}
\end{figure}

Several sources with X-ray luminosities in the range of $42.0<$L$_{X}<43.0$, which show galaxy-like optical spectra, are hard to classify due to the small number of photons in their X-ray spectra. They are marked so far as AGN type 2/Galaxy. Table 1 gives the identification status of the Deep {\em XMM-Newton} Survey.
The majority of the so far spectroscopically identified sources are type 1 and type 2 AGNs. Although we have no complete identification so far we find a strong indication for a larger fraction of type 2 AGNs, especially in the ultra hard sample, compared to that ($\sim$20\%) of the UDS (Lehmann et al. 2001). Nearly all spectroscopically identified type 2 AGNs are at moderate redshift ($z<1$). One type 2 QSO candidate has been identified in the Lockman Hole region so far (see Figure 3). The highest redshift type 2 QSO ($z=3.700$) known to date was recently discovered in the CDF-S survey (Norman et al. 2001).

Figure 4 shows a correlation between the $R$ magnitude and the soft X-ray flux
for {\em ROSAT} and {\em XMM-Newton} sources in the Lockman Hole field. Most sources identified with AGNs follow a correlation line with equal flux in X-ray and optical bands. Nearly half of the unidentified sources (mainly from {\em XMM-Newton}) have very faint 
optical counterparts, which magnitudes are above our current spectroscopic limit ($R=24.0$). As already mentioned, most spectroscopically identified, faint new sources are type 2 AGNs.

The colour-magnitude diagram between the $R$ magnitude and the $R-K'$ colour for 
{\em ROSAT} and {\em XMM-Newton} sources (see Figure 4) shows a clear correlation
between colour, magnitude and AGN type. Type 1 AGNs are relatively bright and blue compared to type 2 AGNs, which show on average a much redder colour. The optical/near-infrared spectra of type 2 AGNs are probably dominated by the light from the host galaxy, while the nucleus
is obscured in the optical/near-infrared wavelength bands (Lehmann et al. 2000).
The optical counterparts of most unidentified faint sources show very red colours ($R-K'>5$), they are extremely red objects (EROs). The {\em XMM-Newton}
spectra of these EROs indicate strong intrinsic absorption (log N$_{H}>22$).

\begin{figure}[t]
\caption{$R-K'$ colour versus redshift for those X-ray sources in 
the Lockman Hole with available $K'$ band photometry. Same symbols are used as in Figure 2. The plus signs give the three EROs with photometric redshifts. The dotted lines correspond to unevolved spectral models for E (upper) and S$_{b}$ (lower) galaxies (Bruzual \& Charlot 1993).
The type 1 AGN showing narrow absorption lines is marked with NALQSO.}
\plotfiddle{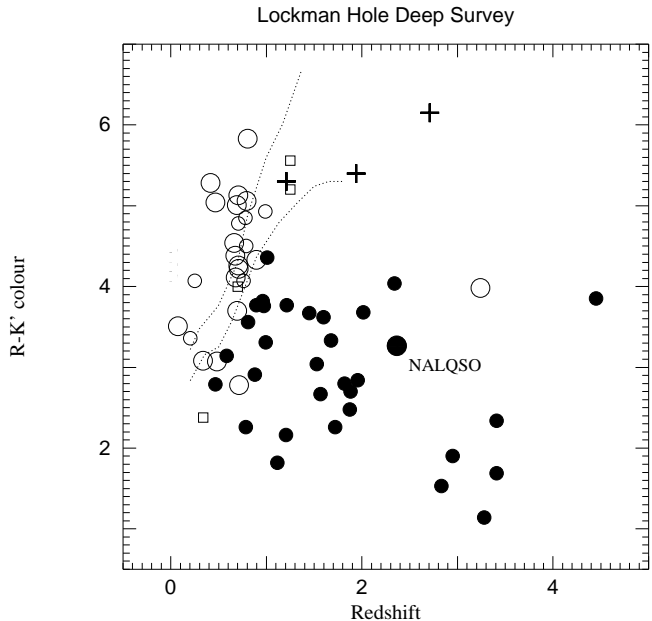}{8cm}{0.0}{50}{50}{-145.0}{0.0}
\end{figure}

For three EROs, already detected in the UDS, we have derived photometric redshifts from
fitting galaxy templates to the spectral energy distribution (SED) of the
objects, obtained from broad-band photometry ($VRIzJHK$). The photometric
redshift spans over the range $1<z<3$ and is based on the assumption that
the SED of these very hard X-ray sources in the optical/near-infrared is due
to stellar processes. The very high intrinsic luminosity (log L$_{X}>44$) may
suggest high redshift examples of luminous obscured AGNs, type 2 QSOs (see
Hasinger 2000, Lehmann et al. 2001). The colour ($R-K'$) versus
redshift diagram (see Figure 5) indicates that nearly all type 2 AGNs are located in
the same region as expected for elliptical and spiral galaxies, which is 
 further evidence that the optical/near-infrared SED of type 2 AGNs is dominated by the light from their host galaxies. 

{\em XMM-Newton} has detected a much
larger number of type 2 AGNs at redshift $z<1$ in the Lockman Hole than previously found with {\em ROSAT}. The number of type 2 AGNs at larger redshifts is still unknown. We have spectroscopically confirmed only one type 2 AGN at $z>1$, a type 2 QSO candidate showing narrow L$\alpha$ and C~III$]$ emission lines redshifted to $z=3.240$. 

Most of the unidentified faint {\em 
XMM-Newton} sources have very faint optical counterparts ($R>24.0$) and at least half of them are extremely red objects (EROs, $R-K'>5.0$). The fraction of
EROs (see Table 1) is still a lower limit, because several X-ray sources are not covered by the deep $K'$ band survey in the Lockman Hole. 
The new {\em XMM-Newton} sources with EROs as optical counterparts are similar to those objects in the UDS with photometric redshifts suggesting obscured AGNs at redshifts $1<z<3$. The photometric redshift technique is probably the
only tool to identify such faint optical objects.
The {\em XMM-Newton} source population at faint fluxes is therefore likely dominated by obscured AGNs (type 2), as predicted by the AGN population synthesis models for the X-ray background. 

\begin{figure}
\caption{Spectral energy distribution (SED) of the ultraluminous IRAS galaxy NGC 6240 from the radio to hard X-ray frequencies (solid line, dashed in unobserved portions), compared to the medium QSO template (thin line) by Elvis et al. (1994). The SED data points of the high-redshift type 1 32A (crosses) and type 2 AGNs 12A and 14Z (filled and open circles) from the {\em ROSAT} Deep Survey are overplotted and have been shifted to the redshift of NGC 6420 and scale by the square of their luminosity distances.  The shape of the cosmic energy density spectrum is indicated by the thick solid and dashed line. The figure is taken from Hasinger et al. 2001b. }
\plotfiddle{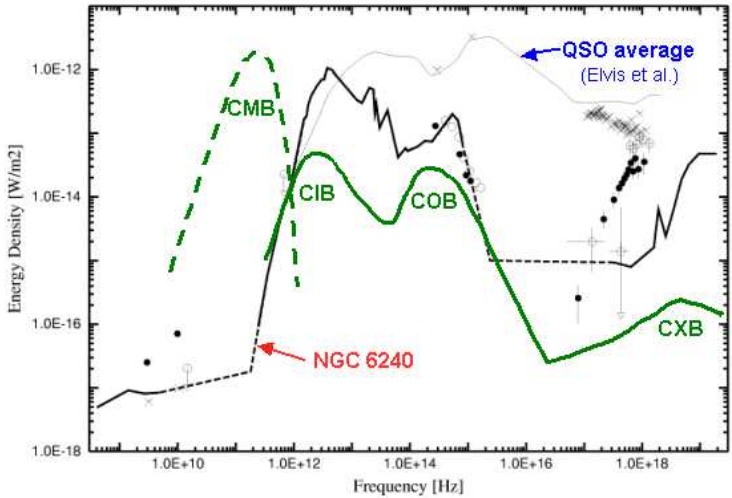}{7cm}{0.0}{50}{50}{-175.0}{-100.0}
\end{figure}

\section{High-z analogues to the obscured AGN in NGC 6420}

The ultraluminous IRAS galaxy NGC 6420 is a key example for a local highly obscured AGN. The object is a classical starburst galaxy with a double nucleus, one of them shows some evidence for a weak activity (Rafanelli et al. 1997). Spectroscopic observations with {\em ISO} have revealed that the far-infrared dust emission is mainly heated by star formation processes (Lutz et al. 1996). Hard X-ray spectroscopy with {\em BeppoSAX} (Vignati et al. 1999) and recently with {\em XMM-Newton} (Keil et al. 2001) discovered a heavily absorbed AGN (N$_{H}\sim10^{24}$ cm$^{-2}$). A significant fraction of the bolometric luminosity of NGC 6420 could
be due to the obscured AGN.

The spectral energy distribution (SED) of NGC 6420, given in Figure 6 from the radio to the hard X-ray band, has a similar shape as the 
cosmic energy density spectrum from the far-infrared to hard X-rays. 
The nonthermal AGN continuum in the hard X-ray range is severely absorbed below 15 keV and strong reprocessed iron line is seen. The UV to soft X-ray AGN continuum is completely obscured. 

In Figure 6 we have superposed the SEDs of two type 2 AGNs (12A at $z=0.99$ and at 14Z $z_{phot}=1.94$) and of the type 1 AGN 32A at $z=1.113$ from the RDS (Schmidt et al. 1998). Both type 2 AGNs show very red colours ($R-K\sim5.0$). The VLA 6cm and 20cm detections and upper limits are taken from Ciliegi et al. (2000) and de Ruiter et al. (1997). The upper limit for the millimeter-range
for 14A was obtained with IRAM (Bertoldi, priv. comm.). The {\em ROSAT} spectra of the sources were fitted with a powerlaw plus galactic or intrinsic absorption. The large intrinsic absorption of the two type 2 AGNs (log N$_{H}\sim22.0$) has been confirmed with {\em XMM-Newton} spectra (V. Mainieri, priv. comm.).

The SED of the two absorbed ROSAT sources (12A and 14Z) is surprisingly 
similar to that of NGC 6420. The higher radio luminosity of 12A could indicate
a relatively strong starburst component. The main difference between the SEDs 
lies in the about a factor of 100 stronger intrinsic absorption of NGC 6240 (log N$_{H}\sim24.0$) compared to 12A and 14Z in the hard X-ray range.
In contrast, the SED of the type 1 AGN 32A is clearly dominated by the unobscured light of the active nucleus in the optical/X-ray bands and is fairly good represented by the averaged QSO SED (see Figure 6.). 

The similarity of the SEDs of obscured AGNs (type 2) and local ultraluminous galaxies may indicate a potential contribution of high-redshift type 2 AGNs to the FIR/submillimeter background. The combination of deep X-ray and submm observation has revealed a first picture about the X-ray/submm correlation of X-ray sources. About 10\% of the X-ray sources seem to be correlated with submm sources (eg., Fabian et al. 2000, Bautz et al. 2000, Hornschemeier et al. 2000, Barger et al. 2001, Almaini et al. 2001), which  agrees quite well with the prediction from Almaini et al. (1999) based on the AGN synthesis models for the X-ray background. Follow-up observations of hard X-ray sources in the mm/sub-mm bands may help to understand the link between the population of obscured AGNs and ultraluminous galaxies in the near future.

\section{Summary}

The soft X-ray sources found with {\em ROSAT} in the Lockman Hole region are mainly X-ray and optically unobscured AGNs (type 1). A significant fraction of obscured AGNs (type 2) has been already detected with {\em ROSAT}. The optical spectra of type 2 AGNs are probably dominated by the light of the host galaxy.

The new {\em XMM-Newton/Chandra} sources detected in the Lockman Hole are harder than the ROSAT sources indicating moderately and highly obscured AGNs (type 2).
The spectroscopic identification of these new sources has revealed large
number of type 2 AGNs at $z<1$. We have spectroscopically identified one high-redshift type 2 QSO candidate ($z=3.240$). 

The {\em XMM-Newton} spectra of type 2 AGNs can be represented by a powerlaw including intrinsic
absorption in the range of N$_{H}=10^{21..24}$ cm$^{-2}$.
Most of the faint so far unidentified hard X-ray sources are optically faint ($R>24$) and are extremely red objects (EROs). These objects could be the high-redshift analogues of heavily obscured AGNs similar to NGC 6240. The photometric redshifts of three very red counterparts of hard X-ray sources indicate such high-redshift heavily obscured type 2 AGNs ($1<z<3$), which may contribute to the FIR/sub-mm background.


\end{document}